\documentclass[aps,pre,twocolumn,showpacs]{revtex4-2}
\usepackage{amsmath,amsfonts,bm, color,amssymb,array}
\usepackage{graphicx,epsfig,overpic,hyperref,ulem}
\usepackage{matlab-prettifier}
\usepackage{verbatim}
\usepackage{comment}
\usepackage{multirow}
\usepackage{soul}
\usepackage[export]{adjustbox}
\usepackage{subcaption}
\usepackage{float}
\usepackage[scr]{rsfso}
\newcommand{\Laplace}{\mathscr{L}}
\newcommand{\eff}{{\mathrm{eff}}}

\newcommand{\colg}[1]{{\color{violet}{#1}}}

\newcommand{\out}{{\mathrm{1}}}
\newcommand{\ins}{{\mathrm{2}}}
\newcommand{\m}{\mathrm{m}}

\begin{document}
\preprint{APS/123-QED}

\title{Diffuse-charge dynamics across a capacitive interface in a DC electric field}

\author{Shuozhen Zhao\textit{$^{1}$}, Bhavya Balu\textit{$^{1}$}, Zongxin Yu\textit{$^{1}$}, Michael J.~Miksis\textit{$^{1}$}, Petia M.~Vlahovska\textit{$^{1}$}}
\affiliation{%
$^{1}$~Engineering Sciences and Applied Mathematics, Northwestern University, Evanston, IL 60208, USA.}

\date{Jan 18th, 2025}

\begin{abstract}

Cells and cellular organelles are encapsulated by nanometrically thin membranes whose main component is a lipid bilayer. In the presence of electric fields, the ion-impermeable lipid bilayer acts as a capacitor and supports a potential difference across the membrane. We analyze the charging dynamics of a planar membrane separating bulk solutions with different electrolyte concentrations upon the application of an applied uniform DC electric field.  The membrane is modeled as a zero-thickness capacitive interface. The evolution of the electric potential and ions distributions in the bulk are solved for using the Poisson-Nernst-Planck (PNP) equations. Asymptotic solutions are derived in the limit of thin Debye layers and weak fields (compared to the thermal electric potential). 

\end{abstract}

\maketitle

\section{Introduction}
An electric potential difference is maintained across most cellular membranes in living cells.
Modulation of the transmembrane electric field 
is a biologically important signal in many cellular functions and physiological processes \cite{Bean:2007,Levine:2021,FUNK:2023}.
The transmembrane potential can be changed intrinsically, through transport of ions across the membrane between the extracellular and intracellular spaces, or externally through application of electric fields. Models for the charging dynamics have been developed, e.g., for the simple geometry of a planar membrane \cite{Vlahovska:APBL, Lacoste:2011}  based on either the Poisson-Nernst-Planck (PNP) equations~\cite{Lacoste:2007, Lomholt:2007,  Lacoste:2009, Ziebert:2010, Lacoste:2010b} or the leaky dielectric model \cite{Vlahovska:2019,Schwalbei:2011, Seiwert:2012,Seiwert:2013}. The latter treats the membrane as a planar capacitor with capacitance $C_\m$ separating fluids with conductivity $\sigma_\ins$ and $\sigma_\out$, and predicts that
the transmembrane potential increases exponentially in time until the entire voltage drop occurs across the membrane
	\begin{equation}
		\label{potentialV}
		 V_\m(t)=   V\left(1-e^{-\frac{t}{t_{\m}}}\right) \,,
	\end{equation}
where
	\begin{equation}
	\label{tm_insul}
t_{\m}=L C_\m \left(\sigma_\ins^{-1}+\sigma_\out^{-1}\right)\,,
\end{equation}
is the capacitor charging time. 
 The result that the charging time is system-size dependent, in this case the distance between the electrodes $L$, seems unphysical, and motivates the present study. In reality, the fluids conductivity is not spatially-uniform since the concentration of the ions near the membrane deviates significantly from the bulk. Upon application of the electric field, the ions brought by conduction accumulate in thin diffuse layer near the membrane physical surfaces \cite{Bazant:2004}. Here, we analyze the evolution of these diffuse layers using the PNP equations. The previous studies \cite{Lacoste:2007, Lomholt:2007,  Lacoste:2009, Ziebert:2010, Lacoste:2010b} have considered only a symmetric membrane, i.e., a membrane separating solutions with the same electrolyte concentrations. Here, we include asymmetry and investigate the impact of the difference in the electrolyte concentration on the transmembrane potential and membrane charging dynamics.

 The paper is organized as follows. Section \ref{Model} defines the model and governing equations. In Section \ref{asymptotic}, we derive an asymptotic solution of both transient and steady state of the time-dependent PNP model in the limit of thin Debye layers and weak electric field. In Sec. \ref{results} we present numerical results for the full problem and compare to the asymptotics.

\section{Problem formulation}\label{Model}
\begin{figure}[b]
    \centering
    \includegraphics[width=0.4\textwidth]{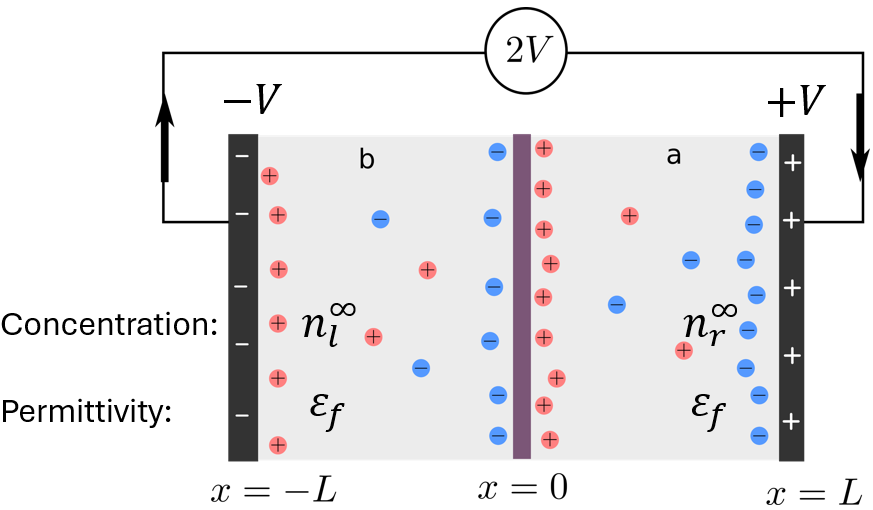}
    \caption{Setup of PNP model. A planar ion-impermeable membrane is at the middle of two electrodes in a simple dilute 1:1 electrolyte with different ion concentration $n_r^\infty\neq n_l^\infty$. A voltage difference $2V$ is applied at $t=0$.}
    \label{fig:pnp}
\end{figure}
We consider a thin planar charge-free membrane placed between two parallel-plate electrodes that are distance $2L$ apart. The problem is sketched in Fig. \ref{fig:pnp}. The membrane separates solutions of 1:1 electrolytes with same permittivity $\epsilon_f$ but different ions concentration $n^\infty_r$ and $n^\infty_l$. In this article, subscripts $r$ and $l$ denote right side $x>0$ and left side $x<0$ of the membrane. At $t = 0$,  a DC voltage with magnitude $2V$ is applied. The response of the ions to the electric field is described by the Nernst-Planck equations:

\begin{align}\label{npeq}
    & \frac{\partial n^{\pm}}{\partial t} + \frac{\partial j^{\pm}}{\partial x} = 0\,,
\end{align}
where in the absence of fluid flow, the flux of ions is
\begin{align}\label{flux-pnp}
    & j^{\pm} = -D\left(\frac{\partial n^{\pm}}{\partial x} \pm \frac{e}{k_B T}n^\pm\frac{\partial\phi}{\partial x} \right)\,.
\end{align}
Here, $D$ is the diffusion coefficient, $k_B T$ is the thermal energy and $\phi$ represents the electric potential. We assume that all ions have the same diffusion coefficient  $D = \omega k_BT$, where $\omega$ is the ion mobility. The relation between the charge density and electric potential is given by the Poisson equation
\begin{align}\label{poisson-pnp}
    & -\varepsilon_f \frac{\partial^2\phi}{\partial x^2} = e(n^+ - n^-)\,.
\end{align}
By introducing the charge density $\rho = e(n^+ - n^-)$ and average concentration $c = n^+ + n^-$, Eqs.~(\ref{npeq})-(\ref{poisson-pnp}) transform into
\begin{subequations}
\begin{align}
    & \frac{\partial\rho}{\partial t} = D\left(\frac{\partial^2\rho}{\partial x^2}+\frac{\partial}{\partial x}\left(\frac{e^2}{k_B T}c\frac{\partial\phi}{\partial x}\right)\right)\,,\label{pnpmodels}\\
    & \frac{\partial c}{\partial t} = D\left(\frac{\partial^2 c}{\partial x^2}+\frac{\partial}{\partial x}\left(\frac{\rho}{k_B T}\frac{\partial}{\partial x}\phi\right)\right)\,,\\
    & -\varepsilon_f\frac{\partial^2 \phi}{\partial x^2} = \rho\,.\label{pnpmodele}
\end{align}
\end{subequations}

Since the focus is on the membrane charging dynamics, the electrode polarization is ignored and we assume that the electrolyte is quasineutral with zero charge density near the electrode~\cite{Ziebert:2010}.  Accordingly, at the electrodes the boundary conditions are
\begin{align}
\begin{split}
    & \rho(x=\pm L) = 0\,,\\
    & c(x=-L) = 2n_l^\infty,\;\;c(x=L) = 2n_r^\infty\,,
\end{split}
\end{align}
and
\begin{align}\label{pnpbc1}
    & \phi(x=\pm L) = \pm V\,.
\end{align}

The membrane is modeled as a zero-thickness interface with capacitance $C_m$. Accordingly, the boundary conditions on the membrane are \cite{Lacoste:2009, Ziebert:2010}
\begin{align}\label{pnpbc2}
    & \left.\varepsilon_f \frac{\partial\phi_l}{\partial x}\right\vert_{x=0} = \left.\varepsilon_f \frac{\partial\phi_r}{\partial x}\right\vert_{x=0} = C_m V_m\,.
\end{align}
$V_m = \phi_r(x=0,t) - \phi_l(x=0,t)$ is the transmembrane potential. The Ampère-Maxwell law \cite{Jackson} in the limit of zero-thickness membrane leads to the current conservation condition at $x=0$:
\begin{align}
    & D\left( \frac{\partial\rho}{\partial x}+\frac{e^2}{k_B T}c \frac{\partial\phi}{\partial x}\right)+\varepsilon_f\frac{\partial}{\partial t}\left( \frac{\partial\phi}{\partial x}\right) = C_m \frac{dV_m}{dt}\,.\label{ztam}
\end{align}
Note that the commonly used zero-flux boundary condition \cite{Bazant:2004} 
\begin{align}
    & \frac{\partial\rho}{\partial x}+\frac{e^2}{k_B T}c \frac{\partial\phi}{\partial x} = 0\,,\label{zeroflux1}
\end{align}
is obtained by taking the time derivatives of Eq.~\eqref{pnpbc2} and substituting them into Eq.~\eqref{ztam}. Since the membrane is impermeable to ions, there is no mass flux of ions through the membrane, and thus
\begin{align}
    &\left.\left(j^+ + j^-\right)\right\vert_{x=0} = \left.\left(\frac{\partial c}{\partial x}+\frac{\rho}{k_B T}\frac{\partial\phi}{\partial x}\right)\right\vert_{x=0} = 0\,.\label{zeroflux2}
\end{align}

\begin{table}[H]
\centering
 \begin{tabular}{|c|c|} 
 \hline
 Unit & Dimensionless variable \\
 \hline\hline
 Concentration &  $\Tilde{c} = c_{r,l}/ (2\sqrt{n_l^\infty n_r^\infty})$\\
 \hline
 Charge density & $\Tilde{\rho} = \rho/(2e\sqrt{n_l^\infty n_r^\infty})$\\
 \hline
 \multirow{2}{4em}{Potential} & $\Tilde{\phi} = \phi/(k_B T/e)$\\
 &$\Tilde{V} = V/(k_B T/e)$\\
 \hline
 Space & $\Tilde{x}=x/L$\\
 \hline
 Time & $\Tilde{t}= t/(\lambda_D L/D)$\\
 \hline
 \end{tabular}
 \caption{Non-dimensionalization for the variables. Dimensionless variables are denoted with tilde $\sim$ on top.}
\label{table:scaling}
\end{table}

\begin{table}[H]
\centering
 \begin{tabular}{|c|c|} 
 \hline
 Parameter & Definition \\
 \hline\hline
 Debye lengths & $\epsilon_{r,l} =\lambda_{r,l}/L$\\
 \hline
 Effective Debye length & $\epsilon =\lambda_D/L$\\
 \hline
 Concentration ratio& $\gamma = \left(n_l^\infty/n_r^\infty\right)^\frac{1}{4}$\\
 \hline
 Conductivity ratio & $R = \sigma_r/\sigma_l = \gamma^{-4}$\\
 \hline
 Membrane capacitance & $\beta = C_m L \epsilon/\varepsilon_f$ \\
 \hline
 Debye capacitance & $C_{r,l} = \varepsilon_f/\lambda_{r,l}$ \\
 \hline
 \end{tabular}
 \caption{Definition of important parameters.}
\label{table:parameters}
\end{table}

\begin{table}[H]
\centering
 \begin{tabular}{|c|c|} 
 \hline
 Physical parameter & Typical value \\
 \hline\hline
 Membrane capacitance $C_m$ & 1 $\mathrm{\mu F/cm^2}$\\
 \hline
 Electrolyte permittivity $\varepsilon_f$ & $80\varepsilon_0$\\
 \hline
 Electrolyte diffusivity $D$ & $10^3\:\mathrm{\mu m^2/s}$\\
 \hline
 Debye lengths $\lambda_D$ & $1\sim 100 \:\mathrm{nm}$\\
 \hline
 Cell length $L$ & $10\sim 100 \:\mathrm{\mu m}$\\
 \hline
 \end{tabular}
 \caption{Typical values of the physical parameters involved in the model.}
\label{table:values}
\end{table}
Hereafter we nondimensionlize all variables using the scales in Table \ref{table:scaling}.
The diffuse layers near the membrane have characteristic widths given by
\begin{align}
    &\lambda_{r,l} = \sqrt{\frac{\varepsilon_f k_B T}{2e^2n_{r,l}^\infty}}\,.
\end{align}
We define an effective Debye length
\begin{align}
    &\lambda_D = \sqrt{\lambda_r\lambda_l}\,.
\end{align}
All important parameters are listed in Table \ref{table:parameters}. The typical values of the physical parameters are listed in Table \ref{table:values}.

To summarize, the dimensionless model and boundary conditions, omitting the tilde over the variables, is
\begin{subequations}
\begin{align}
    & \frac{\partial \rho}{\partial t} = \epsilon\left(\frac{\partial^2 \rho}{\partial x^2} + \frac{\partial}{\partial x}\left(c\frac{\partial \phi}{\partial x}\right)\right)\,,\label{NPs}\\
    &\frac{\partial c}{\partial t} = \epsilon\left(\frac{\partial^2 c}{\partial x^2} + \frac{\partial}{\partial x}\left(\rho\frac{\partial \phi}{\partial x}\right)\right)\,,\label{NPe}\\
    &-\epsilon^2\frac{\partial^2 \phi}{\partial x^2} = \rho\,.\label{dpoi}
\end{align}
\end{subequations}
The boundary conditions at the membrane $x=0$ are
\begin{subequations}
\begin{align}
    &\left.\epsilon\frac{\partial \phi_r}{\partial x}\right\vert_{x=0} = \left.\epsilon\frac{\partial \phi_l}{\partial x}\right\vert_{x=0} = \beta V_m\,,\label{robinbc}\\
    &J = \frac{\partial \rho}{\partial x} + c\frac{\partial \phi}{\partial x} = 0\,,\label{zerofluxs}\\ 
    &\frac{\partial c}{\partial x} + \rho\frac{\partial \phi}{\partial x} = 0\,.\label{zerofluxe}
\end{align}
\end{subequations}
The dimensionless Ampère-Maxwell boundary conditions \eqref{ztam} at $x=0$ is
\begin{align}
    &\left(\frac{\partial \rho_{r,l}}{\partial x}+c_{r,l}\frac{\partial \phi_{r,l}}{\partial x}\right) + \epsilon\frac{\partial}{\partial t}\left(\frac{\partial\phi_{r,l}}{\partial x}\right) = \beta\frac{dV_m}{dt}\,.\label{currentbc}
\end{align}
At the electrode $x=\pm 1$, the boundary conditions are
\begin{subequations}
\begin{align}
    & \phi(x=\pm 1) = \pm V\,,\label{fv}\\
    &\rho(x=\pm 1) = 0\,,\\
    &c(x=-1) = \gamma^2,\quad c(x=1) = \gamma^{-2}\,.\label{bcend}
\end{align}
\end{subequations}
The initial conditions that correspond to zero bulk charge $\rho(t=0)=0$, and are consistent with the zero-thickness condition Eq.~\eqref{robinbc}, are
\begin{subequations}
\begin{align}
    &\phi(t=0) = \begin{cases}
        -V + \frac{V}{1+\frac{\epsilon}{2\beta}}(x+1)& -1<x<0\,,\\
        V+\frac{V}{1+\frac{\epsilon}{2\beta}}(x-1)& 0<x<1\,,
    \end{cases}\label{ICs}\\
    &\rho(t=0) = 0\,,\label{rhoic}\\
    & c(t=0) = \begin{cases}
        \gamma^2 & -1<x<0\,,\\
        \gamma^{-2} & 0<x<1\,.
    \end{cases}\label{ICe}
\end{align}
\end{subequations}

\section{Asymptotic analysis for thin Debye layers}\label{asymptotic}
In this section, we seek an asymptotic solution in the thin Debye layer limit. The limit $\epsilon\rightarrow 0$ is singular, as the solution to Eqs.~\eqref{NPs}-\eqref{dpoi} with $\epsilon=0$ cannot satisfy all boundary conditions (\ref{robinbc})-(\ref{bcend}) with non-zero charge density in the PNP model \cite{Bazant:2004}. Therefore, matched asymptotic expansions are required to obtain the composite solutions.

The initial conditions \eqref{ICs}-\eqref{ICe} do not satisfy the boundary condition \eqref{zerofluxs}. This implies a nonzero initial value of the transmembrane potential $V_m(t=0) = \frac{2V\epsilon}{2\beta+\epsilon} = O(\epsilon)$, which indicates the membrane in our zero-thickness model carries an artificial initial charge $\beta V_m(t=0)$ of order of $\epsilon$. Therefore, the model will experience charge-relaxation dynamics over a time scale $T_\mathrm{cr}$ until the bulk currents  balance at the membrane. This time scale is the bulk charge relaxation time in the leaky dielectric model \cite{Taylor:1969}
\begin{align}
    &T_\mathrm{cr} = \frac{\varepsilon_f}{\sigma}\,,\label{taucre}
\end{align}
suppressing the $r$ and $l$ subscripts. Meanwhile, the bulk currents will charge the membrane and the Debye layers over a time scale $T_\mathrm{mc}$\cite{Bazant:2004}
\begin{align}
    & T_\mathrm{mc} = \frac{C_m L}{\sigma}\,.
\end{align}
The ratio between the two time scales, using the values listed in Table \ref{table:values}, is estimated to be 
\begin{align}
    & \frac{T_\mathrm{mc}}{T_\mathrm{cr}} = \frac{C_m L}{\varepsilon_f} = \frac{\beta}{\epsilon}\approx 200\sim 2000\,,
\end{align}
Thus, the bulk charge-relaxation  is much faster than the membrane-capacitor charging. This implies that these two process are well separated in the overall dynamics.
We carry out the  asymptotic analyses corresponding to these processes in Section~\ref{inisec} and Section~\ref{sec:Membrane-charging dynamics} separately, and in Section~\ref{compasy} we give the composite solutions uniform in time.
 
\subsection{Charge-relaxation dynamics}\label{inisec}
The formal asymptotics for the charge-relaxation dynamics is solved by matched asymptotics in space and Laplace transform in time. Details are provided in Appendix \ref{appini}. Here, we present a physically intuitive approach to solve the charge-relaxation dynamics. The dimensionless charge-relaxation scale is $\Tilde{T}_\mathrm{cr} = \epsilon$ after tentatively setting $\sigma = \sqrt{\sigma_r\sigma_l}$. Re-scaling time as $T = \frac{t}{\epsilon}$, the leading order of the PNP equations \eqref{NPs}-\eqref{dpoi} results in
\begin{subequations}
\begin{align}
    & \rho = 0\,,\label{bulks}\\
    & c = \begin{cases}
        \gamma^2 & -1<x<0\,,\\
        \gamma^{-2} & 0<x<1\,.
    \end{cases}\label{bulkm}
\end{align}
\end{subequations}
Since the membrane is not charging during $T_\mathrm{cr}$ at leading order and the initial transmembrane potential is of order $\epsilon$, we assume there is no inner region during the $T_\mathrm{cr}$ and look for solutions valid in the whole space. Therefore, the leading-order potential is still linear and continuous across the membrane:
\begin{subequations}
\begin{align}
    & V_m = 0\,,\\
    & \phi = \begin{cases}
        -V+E_l(T)(x+1) & -1<x<0\,,\\
        V+E_r(T)(x-1) & 0<x<1\,,
    \end{cases}\label{bulke}
\end{align}
\end{subequations}
where $E_{r,l}$ are unknowns. Additionally, from the left side of Eq.~\eqref{currentbc} we have

\begin{align}
    &\frac{\partial \rho_l}{\partial x}+c_l\frac{\partial \phi_l}{\partial x}+\frac{\partial^2 \phi_l}{\partial T\partial x} = \frac{\partial \rho_r}{\partial x}+c_r\frac{\partial \phi_r}{\partial x}+\frac{\partial^2 \phi_r}{\partial T\partial x}\,.\label{ele}
\end{align}
Eq.~\eqref{ele} is equivalent to the surface charge relaxation dynamics in the leaky dielectric model~\cite{Taylor:1969}:
\begin{align}
    &\frac{dQ_{\mathrm{membrane}}}{dT}+ {\colg{\epsilon}}\left(J_r-J_l\right) = 0\,,\label{leakyini}
\end{align}
where $Q_{\mathrm{membrane}} ={\colg{\epsilon}}\left(E_r-E_l\right)$ is defined as the imbalance of the bulk charge density on the two sides of the membrane~\cite{Schwalbei:2011}. Solving Eqs.~\eqref{bulks}-\eqref{bulke} and \eqref{ele} with the leading order of initial conditions \eqref{ICs}-\eqref{ICe}, results in
\begin{subequations}
\begin{align}
    & \tau_b = \frac{2\epsilon}{\gamma^2+\gamma^{-2}}\,, \label{taucr}\\
    & E_l(T) = \frac{2V}{1+\gamma^4}\left( 1+\frac{\gamma^4-1}{2}e^{-\frac{T}{\tau_b/\epsilon}} \right)\,,\label{sds}\\
    & E_r(T) = \frac{2\gamma^4 V}{1+\gamma^4}\left( 1+\frac{\gamma^{-4}-1}{2}e^{-\frac{T}{\tau_b/\epsilon}} \right)\,.\label{sde}
\end{align}
\end{subequations}

Comparing Eq.~\ref{taucre} and Eq.~\eqref{taucr} shows the effective conductivity in Eq.~\eqref{taucre} is the average of two bulk fluid conductivities $\sigma = \frac{\sigma_r+\sigma_l}{2}$. 

\subsection{Membrane-charging dynamics}\label{sec:Membrane-charging dynamics}
The short-time charge-relaxation dynamics at $T\rightarrow\infty$, provides the initial conditions for the membrane-charging dynamics in the outer region $\phi(T\rightarrow\infty) = \phi(t=0)$:
\begin{subequations}
\begin{align}
    & \phi(t=0) = \begin{cases}
        -V + \frac{2V}{1+\gamma^2}(x+1) & -1<x<0\,,\\
         V+\frac{2\gamma^2V}{1+\gamma^2}(x-1) & 0<x<1\,,
    \end{cases}\\\label{longICs}
    &\rho(t=0)=0\,,\\
    & c(t=0) = \begin{cases}
        \gamma^2 & -1<x<0\,,\\
        \gamma^{-2} & 0<x<1\,,
    \end{cases}\label{longICe}
\end{align}
\end{subequations}
\subsubsection{Nonlinear dynamics with finite voltage}\label{nlasy}
Although shortly we will consider the weak-field assumption, we begin our analysis of the membrane-charging dynamics by assuming $V=O(1)$. A more rigorous matched asymptotic analysis in space is required to proceed at this stage. The outer ``bulk" region is studied using a regular asymptotic expansion. The Poisson equation~\eqref{dpoi} indicates that the charge density $\rho$ is at most second order:
\begin{subequations}
\begin{align}
    & c = c^{(0)} + \epsilon c^{(1)} + \dots\,,\label{asys}\\
    & \rho = \epsilon^2 \rho^{(2)} + \epsilon^3 \rho^{(3)} + \dots\quad (\rho^{(0)} = \rho^{(1)} = 0)\,,\\
    & \phi = \phi^{(0)} + \epsilon \phi^{(1)}+\dots\,.\label{asye}
\end{align}
\end{subequations}
Substituting expansions~\eqref{asys}-\eqref{asye} into Eqs.~\eqref{NPs}-\eqref{dpoi} eliminates the time dependence at leading order. Eq.~\eqref{NPe} and Eq.~\eqref{NPs} indicate that the  concentration $c^{(0)}$ is constant and the potential $\phi^{(0)}$ is linear in space:
\begin{subequations}
\begin{align}
    & c^{(0)} =\begin{cases}
        \gamma^2 & -1<x<0\,,\\
        \gamma^{-2} & 0<x<1\,,
    \end{cases}\label{outer}\\
    & \phi^{(0)} = \begin{cases}
        \gamma^{-2}j_l(t)x+b_l(t) & -1<x<0\,,\\
        \gamma^2j_r(t)x+b_r(t) & 0<x<1\,,
    \end{cases}\label{outerphi}
\end{align}
\end{subequations}
where the bulk currents $j_{r,l}(t)$, and $b_{r,l}(t)$ are unknown. Eq.~\eqref{dpoi} shows the leading order bulk charge density $\rho^{(2)} = -\frac{\partial^2 \phi^{(0)}}{\partial x^2}$ is also zero,
\begin{subequations}
\begin{align}
    & \rho^{(2)} = 0\,.
\end{align}
\end{subequations}

The outer region expansions are matched with inner layer expansions  in the Debye layers. Variables $C$, $P$ and $\Phi$ are used to denote concentration, density and potential inside the Debye layers. Reformulations of Eqs.~\eqref{NPs}-\eqref{dpoi} using the inner coordinates near the membrane $Y = \frac{x}{\epsilon}$ yield
\begin{subequations}
\begin{align}
    &\epsilon\frac{\partial P}{\partial t} = \left(\frac{\partial^2 P}{\partial Y^2} + \frac{\partial}{\partial Y}\left(C\frac{\partial \Phi}{\partial Y}\right)\right)\,,\label{inners}\\
    &\epsilon\frac{\partial C}{\partial t} = \left(\frac{\partial^2 C}{\partial Y^2} + \frac{\partial}{\partial Y}\left(P\frac{\partial \Phi}{\partial Y}\right)\right)\,,\label{innerm}\\
    &-\frac{\partial^2 \Phi}{\partial Y^2} = P\,.\label{innere}
\end{align}
\end{subequations}

$C$, $P$, $\Phi$ are expanded as follows:

\begin{subequations}
\begin{align}
    &C(Y,t) = C^{(0)} + \epsilon C^{(1)}\label{asyinners} + \cdots\,, \\
    &P(Y,t) = P^{(0)} + \epsilon P^{(1)} + \cdots\,, \\
    &\Phi(Y,t) = \Phi^{(0)} + \epsilon \Phi^{(1)} + \cdots\,. \label{asyinnere}
\end{align}
\end{subequations}
The boundary conditions are obtained from matching $Y \rightarrow \pm\infty$ to $x=0$:
\begin{subequations}
\begin{align}
    &C_l^{(0)}(-\infty,t) = \gamma^2,\;C_r^{(0)}(\infty,t) = \gamma^{-2}\,,\\
    &P(\pm\infty,t) = 0\,,\\
    &\Phi_l^{(0)}(-\infty,t) = \phi_l^{(0)}(0,t) = b_l(t)\,,\\
    &\Phi_r^{(0)}(\infty,t) = \phi_r^{(0)}(0,t) = b_r(t)\,.
\end{align}
\end{subequations}
After substituting expansions \eqref{asyinners}-\eqref{asyinnere} into the inner equations \eqref{inners}-\eqref{innere}, again, the time dependency vanishes at leading order. Utilizing the zero-flux boundary conditions \eqref{zerofluxs}-\eqref{zerofluxe},  the leading order inner equations inside the Debye layers become
\begin{subequations}
\begin{align}
    & \frac{\partial P^{(0)}}{\partial Y} + C^{(0)}\frac{\partial \Phi^{(0)}}{\partial Y} = 0\,,\label{mi1}\\
    & \frac{\partial C^{(0)}}{\partial Y} + P^{(0)}\frac{\partial \Phi^{(0)}}{\partial Y} = 0\,,\label{mi2}\\
    &-\frac{\partial^2 \Phi^{(0)}}{\partial Y^2} = P^{(0)}\,.\label{mipo}
\end{align}
\end{subequations}
Eqs.~\eqref{mi1}-\eqref{mi2} give the classical Gouy-Chapman profiles for the concentration and charge density. For convenience, excess voltages are defined as
\begin{subequations}
\begin{align}
    & \Psi^{(0)}(Y,t) = \Phi^{(0)}(Y,t) - \phi^{(0)}(0,t)\,,\label{excessv}\\
    &\Psi^{(0)}(\pm\infty,t) = 0\,.\label{matchex}
\end{align}
\end{subequations}
Eq.~\eqref{matchex} follows from the matching conditions. The solutions to Eqs.~\eqref{mi1}-\eqref{mi2} are
\begin{align}\label{mis}
\begin{split}
    & C_l^{(0)} = \gamma^2\cosh(\Psi_l^{(0)}),\quad C_r^{(0)} = \gamma^{-2}\cosh(\Psi_r^{(0)})\,,\\
    & P_l^{(0)} = -\gamma^2\sinh(\Psi_l^{(0)}),\quad P_r^{(0)} = -\gamma^{-2}\sinh(\Psi_r^{(0)})\,.
\end{split}
\end{align}
Substituting Eqs.~\eqref{mis} into the Poisson equation~\eqref{mipo}, and solving the resulting equations subject to the matching conditions \eqref{matchex}, yields

\begin{subequations}
\begin{align}\label{innerphiNon}
    & \Psi^{(0)} = \begin{cases}
        4\tanh^{-1}(A_l^{(0)}(t)e^{\gamma Y}/4) & -\infty<Y<0\,,\\
        4 \tanh^{-1}(A_r^{(0)}(t)e^{-\frac{Y}{\gamma}}/4) & 0<Y<\infty\,,
    \end{cases}\\
    & A_l^{(0)}(0) = A_r^{(0)}(0) = 0\label{amic}\,,
\end{align}
\end{subequations}
where $A_l^{(0)}(t)$ and $A_r^{(0)}(t)$ are integration constants to be determined by the boundary conditions at $Y=0$.  Eqs.~\eqref{innerphiNon}-\eqref{amic} along with Eqs.~\eqref{mis} describe the leading order behavior of the inner solution. To obtain an explicit analytical solution, we assume that the applied voltage is small, $V\ll 1$, which enables us to linearize the solution. This calculation is conducted in the next subsection.

\subsubsection{Linear dynamics with small voltage}\label{lasy}
The small voltage assumption $V\ll 1$ implies that $\lvert A^{(0)}(t)\rvert\ll 1$, since $\lvert A^{(0)}(t)\rvert$ is linear with $V$. This will become evident at the end of this subsection in Eqs.~\eqref{tt1}-\eqref{tt2}. Expanding Eq.~\eqref{innerphiNon} with small $\lvert A^{(0)}(t)\rvert$ yields the leading order linearized potentials

\begin{subequations}
\begin{align}\label{innerphi}
    & \Psi_l^{(0)} = A_l^{(0)}(t)e^{\gamma Y}+O(\lvert A_l^{(0)}(t)\rvert^3)\,,\\
    & \label{innerphi2} \Psi_r^{(0)} = A_r^{(0)}(t)e^{-\frac{Y}{\gamma}}+O(\lvert A_r^{(0)}(t)\rvert^3)\,.
\end{align}
\end{subequations}
Substituting \eqref{innerphi}-\eqref{innerphi2} into the Robin-type boundary conditions \eqref{robinbc} at the membrane results in
\begin{align}
\begin{split}
    \gamma A_l^{(0)}(t) & = -\gamma^{-1}A_r^{(0)}(t)\\
    & =\beta\left(A_l^{(0)}(t)-A_l^{(0)}(t)+b_r(t)-b_l(t)\right)\\
    & =\beta V_m^{(0)}(t)\,.\label{robinasy}
\end{split}
\end{align}
In order to find the explicit time dependence of $A_l^{(0)}$ and $A_r^{(0)}$, we can either do a full asymptotic analysis using Laplace transform, similar to the one described in Appendix \ref{appini}, or follow the approach of Ref. \cite{Bazant:2004}, which considers the evolution of the diffuse charge $Q$ within the Debye layers, where $Q$ is defined as the integral of the charge density inside the Debye layers. The leading order of $Q$ is obtained from the inner Poisson equation \eqref{mipo}, as
\begin{subequations}\label{innercharge}
\begin{align}
\begin{split}
    &Q_l^{(0)}(t) = \int_{-\infty}^0 P_l^{(0)}(Y,t)dY = -\left.\frac{\partial\Phi_l^{(0)}}{\partial Y}\right\vert_{-\infty}^0\\
    \Rightarrow\:& Q_l^{(0)}(t)=-\gamma A_l^{(0)}(t)\,,
\end{split}
\end{align}
Similarly,
\begin{align}
    & Q_r^{(0)}(t)  = -\gamma^{-1}A_r^{(0)}(t)\,. \label{Qr0}
\end{align}
\end{subequations}
The time derivatives of $Q_{r,l}^{(0)}$ are
\begin{subequations}
\begin{align}
    & \frac{dQ_l^{(0)}}{dt} = -\gamma\frac{dA_l^{(0)}}{dt}\,,\label{dal}\\
    & \frac{dQ_r^{(0)}}{dt} = -\gamma^{-1}\frac{dA_r^{(0)}}{dt}\,.\label{dar}
\end{align}
\end{subequations}
The zero-flux boundary conditions \eqref{zerofluxs}-\eqref{zerofluxe} and the matching to derivatives (outer flux) are utilized along with the inner Nernst-Planck equations \eqref{inners}-\eqref{innerm}:
\begin{align}\label{chargederivative}
\begin{split}
    \frac{dQ_l}{dt} & = \int_{-\infty}^0 \frac{\partial P_l}{\partial t}dY\\
    & = \lim_{Y\rightarrow-\infty}-\frac{1}{\epsilon}\left(\frac{\partial P_{l}}{\partial Y}+C_{l}\frac{\partial \Psi_{l}}{\partial Y}\right)\\
    & = \lim_{x\rightarrow 0}-\left(\frac{\partial \rho_{l}}{\partial x}+c_{l}\frac{\partial \phi_{l}}{\partial x}\right)\,.
\end{split}
\end{align}
Next, we consider the leading order of \eqref{chargederivative}:
\begin{subequations}
\begin{align}
\begin{split}\label{dql}
    \frac{dQ_l^{(0)}}{dt} & = \lim_{x\rightarrow 0}-\left(\frac{\partial \rho_l^{(0)}}{\partial x}+c_l^{(0)}\frac{\partial \phi_l^{(0)}}{\partial x}\right)\\
    & =-j_l(t)\,,
\end{split}
\end{align}
Similarly,
\begin{align}
    \frac{dQ_r^{(0)}}{dt} & = j_r(t)\,.\label{dqr}
\end{align}
\end{subequations}
Finally, matching Eqs.~\eqref{dql}-\eqref{dqr} with Eqs.~\eqref{dal}-\eqref{dar} gives rise to the relation between $A^{(0)}$ and $j(t)$:
\begin{subequations}
\begin{align}
    & \gamma\frac{dA_l^{(0)}}{dt}= j_l(t)\,,\label{dynamics}\\
    & -\gamma^{-1}\frac{dA_r^{(0)}}{dt}= j_r(t)\,.\label{dynamice}
\end{align}
\end{subequations}
Substituting \eqref{dynamics}-\eqref{dynamice} into Eqs.~\eqref{robinasy} results in
\begin{align}\label{jrl}
    & j_l(t) = j_r(t):=j(t)\,.
\end{align}
Eq~\eqref{jrl} shows the bulk currents are continuous, so there is no net charge accumulated on the membrane at leading order. It also indicates that the appropriate initial condition for the $O(1)$ time scale is indeed the ``long-time" initial conditions~\eqref{longICs}-\eqref{longICe}~\cite{Bazant:2004}.  Hence, the initial conditions for $j_{r,l}(t)$ are
\begin{subequations}
\begin{align}
    & j(0) = \frac{2\gamma^2V}{1+\gamma^4}\,,\label{inics}\\
    & b_l(0) = b_r(0) = \frac{(1-\gamma^2)V}{1+\gamma^2}\,.\label{inice}
\end{align}
\end{subequations}
Eq.~\eqref{inice} comes from the fixed voltage boundary condition \eqref{fv}:
\begin{subequations}
\begin{align}
    & -V = -j(t)+b_l(t)\,,\label{fvbcs}\\
    & V = j(t)+b_r(t)\,.\label{fvbce}
\end{align}
\end{subequations}
The explicit membrane-charging asymptotics are solved from equations \eqref{fvbcs}-\eqref{fvbce}, \eqref{amic}-\eqref{robinasy},\eqref{dynamics}-\eqref{dynamice} and \eqref{inics}-\eqref{inice}. The membrane-charging time scale is
\begin{align}
    & \tau_m = \frac{\beta(\gamma^2 + \gamma^{-2})}{1+\beta(\gamma^2 + \gamma^{-2})}\,,\label{tau}
\end{align}
and other tentative results are
\begin{subequations}
\begin{align}
     j(t) &= \frac{2V}{\gamma^2+\gamma^{-2}}e^{-\frac{t}{\tau_m}}\,,\\
     A_l^{(0)}(t) &= \frac{2\beta\gamma^{-1} V}{1+\beta(\gamma+\gamma^{-1})}\left(1-e^{-\frac{t}{\tau_m}}\right)\,,\label{tt1}\\
     A_r^{(0)}(t) &= -\frac{2\beta\gamma V}{1+\beta(\gamma+\gamma^{-1})}\left(1-e^{-\frac{t}{\tau_m}}\right)\,.\label{tt2}
\end{align}
\end{subequations}
The full set of composite solutions is summarized in the next section.
\subsection{Full composite asymptotic solution}\label{compasy}
For convenience, from Eqs.~\ref{sds}-\ref{sde} we define
\begin{align}
    & e(T) = V\frac{\gamma^2-\gamma^{-2}}{\gamma^2+\gamma^{-2}}e^{-\frac{T}{\tau_b/\epsilon}}\,.
\end{align}
The leading order composite solution is obtained by combining the charge-relaxation and membrane-charging asymptotic results as
\begin{subequations}
\begin{align}
    & \phi_{lc} = -V+\left[\gamma^{-2}j(t)+e(T)\right](x+1)+A_l^{(0)}(t)e^{\frac{\gamma x}{\epsilon}}\,,\label{philc}\\
    & \phi_{rc} = V+\left[\gamma^2j(t)-e(T)\right](x-1)+A_r^{(0)}(t)e^{-\frac{x}{\gamma\epsilon}}\,,\label{phirc}\\
    & \rho_c = \begin{cases}
        -\gamma^2A_l^{(0)}(t)e^{\frac{\gamma x}{\epsilon}} & -1<x<0\,,\\
        -\gamma^{-2}A_r^{(0)}(t)e^{-\frac{x}{\gamma\epsilon}} & 0<x<1\,,
    \end{cases}\label{rhoc}\\
    & c_c = \begin{cases}
        \gamma^2 +\frac{\gamma^2}{2}\lvert A_l^{(0)}(t)\rvert^2e^{\frac{2\gamma x}{\epsilon}} & -1<x<0\,,\\
        \gamma^{-2} + \frac{\gamma^{-2}}{2}\lvert A_r^{(0)}(t)\rvert^2e^{-\frac{2x}{\gamma\epsilon}} & 0<x<1\,,
    \end{cases}\label{cc}
\end{align}
\end{subequations}
with subscript $c$ indicating composite solution. The error Eqs.~\eqref{philc}-\eqref{cc} is of order $O(\epsilon)+O(V^3)$. Note that substituting Eqs.~\eqref{philc}-\eqref{cc} to Eq. \eqref{zerofluxs} does not yield the correct composite solution for the current $J$, as the bulk current matches one order lower of inner current. It necessitates investigating the second order of inner PNP equation (details are provided in Appendix~\ref{appcurrent}), to obtain the corrected composite current solutions 
\begin{align}
& J_c = \begin{cases}
    j(t)\left(1-e^{\gamma \frac{x}{\epsilon}}\right)+\gamma^2e(T) & -1<x<0\,,\\
    j(t)\left(1-e^{-\frac{x}{\gamma\epsilon}}\right)-\gamma^{-2}e(T) & 0<x<1\,.
\end{cases}\label{compcurr}
\end{align}
The composite solutions lead to the potential jump at the membrane (transmembrane potential) $V_m = \phi_{rc}(x=0) - \phi_{lc}(x=0)$ and the bulk region potential jump $V^o_m = \phi_r^{(0)}(x=0) - \phi_l^{(0)}(x=0)$, with subscript $o$ indicating outer,
\begin{subequations}
\begin{align}\label{vmvmo}
\begin{split}
    V_m &=V\frac{2\gamma + 2\beta\gamma^2 e^{-1/(\gamma\epsilon)}+2\beta e^{-\gamma/\epsilon}}{\gamma+\beta+\beta\gamma^2}\left(1-e^{-\frac{t}{\tau_m}}\right)\\
    &\approx \frac{2 V\left(1-e^{-\frac{t}{\tau_m}}\right)}{1+\beta(\gamma+\gamma^{-1})}\,,
\end{split}\\
    V^o_m &= 2V\left(1-e^{-\frac{t}{\tau_m}}\right)\,.
\end{align}
\end{subequations}
Since the bulk charge density is zero at leading order, we define the composite diffuse charge $Q^m$ within the Debye Layers as the integral of the charge density from the midpoint to the membrane, i.e., $Q^m_l = \int_{-1/2}^0\rho_cdx$ and $Q^m_r = \int_0^{1/2}\rho_cdx$.  Using Eqs.~\eqref{rhoc} we find
\begin{align}
\begin{split}
    Q^m_l = -Q^m_r& =-V\frac{2\epsilon\beta\gamma \left(1-e^{-\frac{t}{\tau_m}}\right)}{\gamma+\beta+\beta\gamma^2}\left(1-e^{-\frac{\gamma}{2\epsilon}}\right)\\
    & \approx -V\frac{2\epsilon\beta \left(1-e^{-\frac{t}{\tau_m}}\right)}{1+\beta(\gamma+\gamma^{-1})}\,,\label{Qmlc}
\end{split}
\end{align}
This indicates the net diffuse charge, $Q^m_c = Q^{m}_l+Q^{m}_r$, inside the Debye layers is zero at leading order, 
\begin{align}\label{neutral}
    Q^{m(0)}_c = Q^{m}_l+Q^{m}_r = 0\,.
\end{align}
In fact, by substituting the derivatives of Eqs.~\eqref{innerphiNon} into Eqs.~\eqref{innercharge}-\eqref{Qr0} and \eqref{robinbc}, this conclusion \eqref{neutral} also holds true in the nonlinear regime $V = O(1)$. Additionally, Eqs.~\eqref{robinasy}-\eqref{Qr0} shows $Q^m_{r,l}$ is linear with $V_m$:
\begin{align}
    & Q^m_r = -Q^m_l = \epsilon\beta V_m\;.\label{qmv}
\end{align}
The nonzero correction to the net diffuse charge appears at the order of $O(\epsilon^2)$ (details in Appendix~\ref{appini} and \ref{appcurrent}), hence the composite solution of net diffuse charge up to the order of $O(\epsilon^2)$ is
\begin{align}
    & Q^m_c= 2\epsilon^2 V\frac{1-\gamma^4}{1+\gamma^4}\left(e^{-\frac{t}{\tau_m}}-e^{-\frac{T}{\tau_b/\epsilon}}\right)\,.\label{totaldiffusecharge}
\end{align}
To get the full composite solutions starting from the ``long-time" initial conditions \eqref{longICs}-\eqref{longICe}, one simply removes terms involving $\tau_b$ in Eqs.~\eqref{philc}-\eqref{compcurr}.

The steady state solutions are obtained by pushing $t\rightarrow\infty$ in Eqs.~\eqref{philc}-\eqref{Qmlc}. Note that setting $\gamma = 1$, the steady state recovers the same solution as in Ref.~\cite{Lomholt:2007}, with the zero-thickness membrane and symmetric dielectric constants on each side, and as in Ref.~\cite{Ziebert:2010} with zero flux at membrane.

\begin{figure*}[t]
\centering
\includegraphics[width=\textwidth]{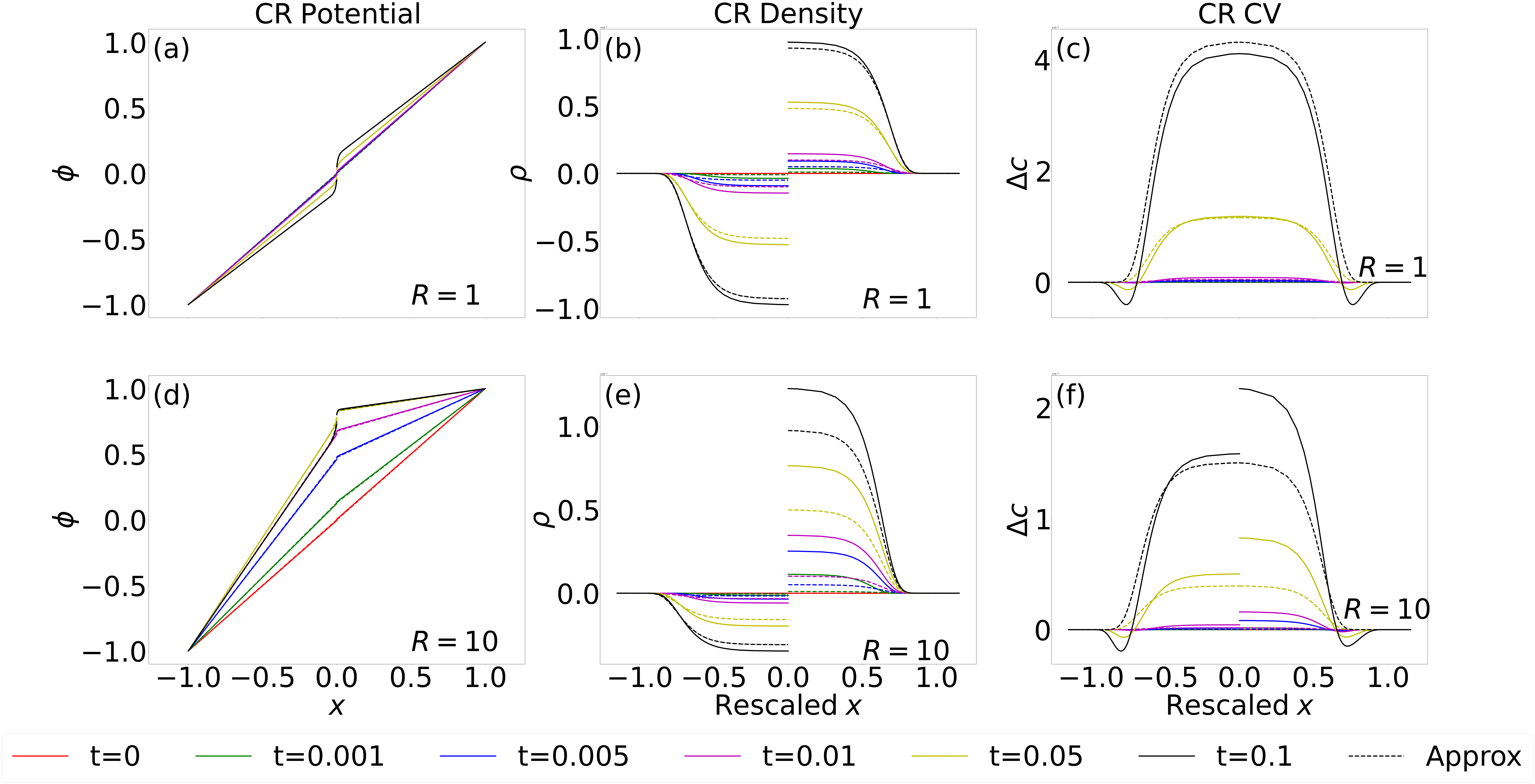}
\caption{Numerical (solid line) and asymptotic (dashed line) transient states of the short-time charge-relaxation (CR) potential $\phi$, density $\rho$ and concentration variation (CV) $\Delta c$ with $R = 1,10$ at $t=0\;(\mathrm{red})$, $0.001\;(\mathrm{green})$, $0.005\;(\mathrm{blue})$, $0.01\;(\mathrm{pink})$, $0.05\;(\mathrm{yellow})$, $0.1\;(\mathrm{black})$ in $O(\epsilon)$ time scale. The colors of dashed line indicate the same time mark as the colors of solid line. The rescaled $x$ represents the mapping to the function $\mathrm{sgn}(x)(\ln{(10^5\lvert x\rvert)+1})/10$, which sets membrane at $0$ and electrodes at $\pm 1$. The parameter setup is $\beta = 1,\epsilon=0.01, V=1$.}
\label{fig:CR}
\end{figure*}

\subsection{Relation between the PNP and Leaky Dielectric Model solutions}
The \textit{dimensional} membrane-charging time scale is
\begin{align}
    & \tau^*_m = \tau_c \tau_m = \frac{L}{\frac{1}{C_m}+\frac{1}{C_r}+\frac{1}{C_l}}\left(\sigma_r^{-1}+\sigma_l^{-1}\right)\,.
\end{align}
If we define an \textit{effective} capacitance,
\begin{align}
    & \frac{1}{C_{\eff}} = \frac{1}{C_m}+\frac{1}{C_l}+\frac{1}{C_r}\label{effective},
\end{align}
then the \textit{dimensional} time scale $\tau^*_m$ of PNP model has the same form as in leaky dielectric model Eq.~\eqref{tm_insul}:
\begin{align}
    & \tau^*_m = C_{\eff}L\left(\sigma_r^{-1}+\sigma_l^{-1}\right)\,.
\end{align}
This inspires us to examine the dynamics of the bulk region. Eq.~\eqref{outerphi} shows the \textit{dimensional} bulk potential satisfies the Laplace equation:
\begin{align}
    & \frac{\partial^2 \phi^{(0)}}{\partial x^2} = 0\,.\label{bulkds}
\end{align}
Taking the time derivatives of Eq.~\eqref{robinasy}, substituting Eqs.~\eqref{dynamics}-\eqref{jrl} and then transforming back to \textit{dimensional} variables leads to
\begin{align}
    & \left.\sigma_l\frac{\partial \phi_l^{(0)}}{\partial x}\right\vert_{x=0} = \left.\sigma_r\frac{\partial \phi_r^{(0)}}{\partial x}\right\vert_{x=0} = C_{\eff}\frac{\partial V^o_m}{\partial t}\,.\label{bulkde}
\end{align}
Thus, the bulk region dynamics~\eqref{bulkds}-\eqref{bulkde} from the PNP model is exactly the same as the leaky dielectric model~\cite{Schwalbei:2011}. Therefore, at leading order, if we combine two Debye layers and the membranne as a single capacitance, the PNP model is equivalent to the effective leaky-dielectric bulk region plus two Debye layers. This conclusion recovers the same result as Song \textit{et al.} (2018) \cite{Zilong:2018} with linearization.

\begin{figure*}[t]
\centering
\includegraphics[width=\textwidth]{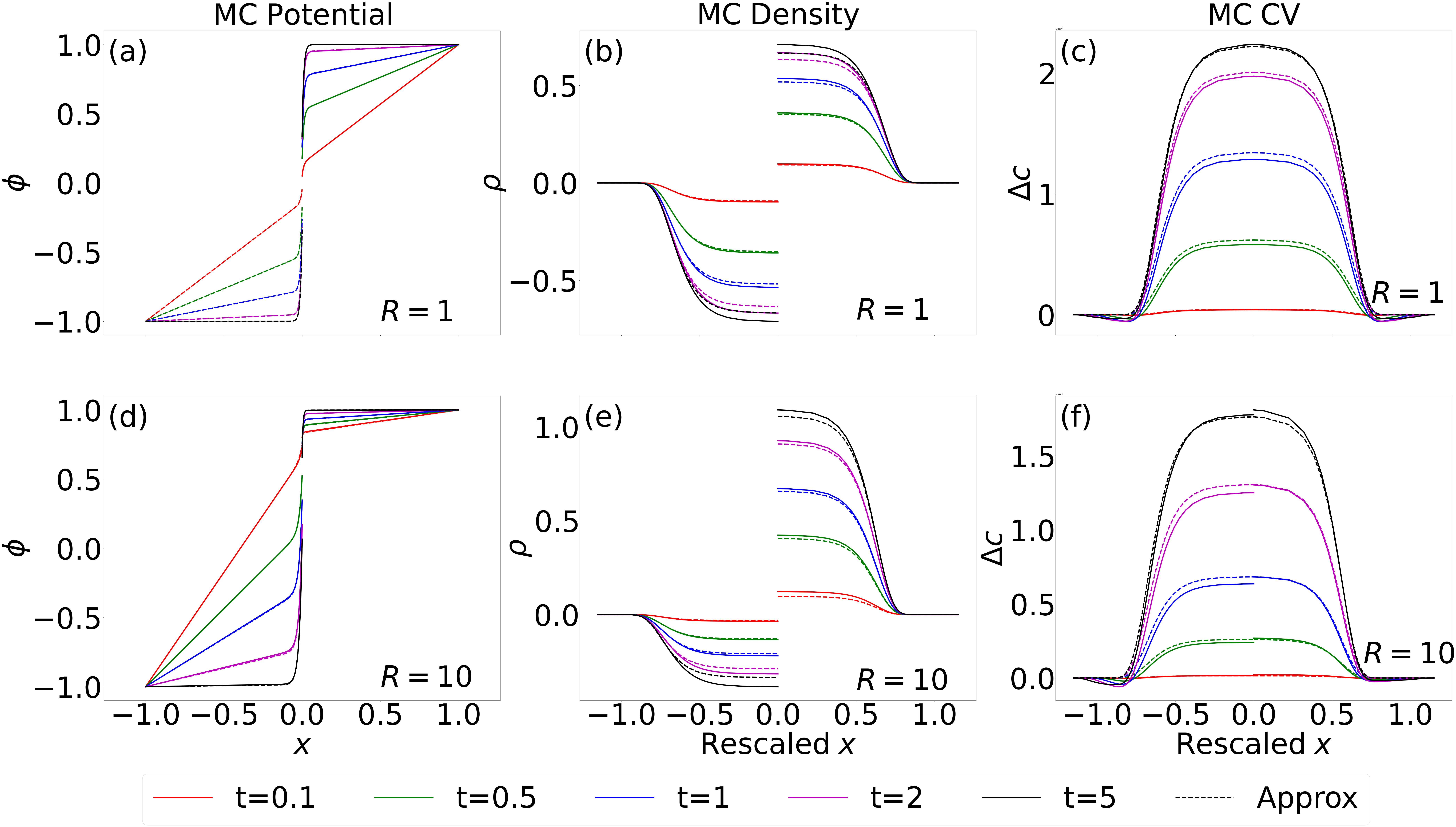}
\caption{Numerical (solid line) and asymptotic (dashed line) transient states of the membrane-charging (MC) potential $\phi$, density $\rho$ and concentration variation (CV) $\Delta c$ with $R = 1,10$ at $t=0.1\;(\mathrm{red})$, $0.5\;(\mathrm{green})$, $0.1\;(\mathrm{blue})$, $2\;(\mathrm{pink})$, $5\;(\mathrm{black})$ in $O(1)$ time scale. The colors of dashed line indicate the same time mark as the colors of solid line. The rescaled x is the same as in Fig.~\ref{fig:CR}. The parameter setup is $\beta = 1,\epsilon=0.01, V=1$}
\label{fig:MC}
\end{figure*}

\begin{figure*}[t]
\centering
\includegraphics[width=\textwidth]{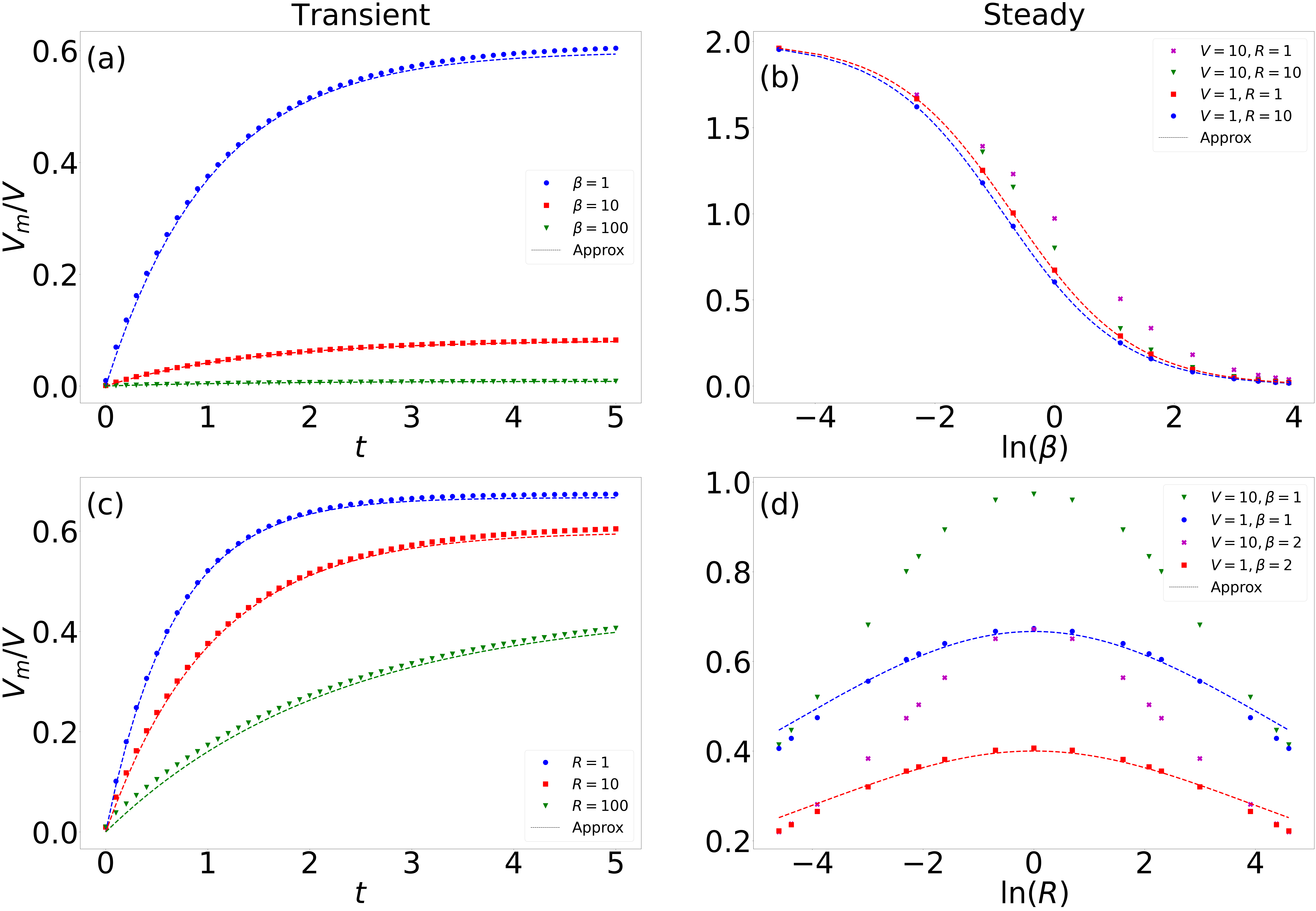}
\caption{Numerical (markers) and asymptotic (dashed line) distributions of transient transmembrane potential $V_m$ over time ($R=10,V=1$) and steady-state $V_m$ over $\beta$ ($R=10$) and $R$ ($\beta=1$) in log scale with imposed voltage $V=1$ and $V=10$. The colors of dashed line have the same legend as the colors of markers.}
\label{fig:vm}
\end{figure*}

\begin{figure*}[t]
\centering
\includegraphics[width=\textwidth]{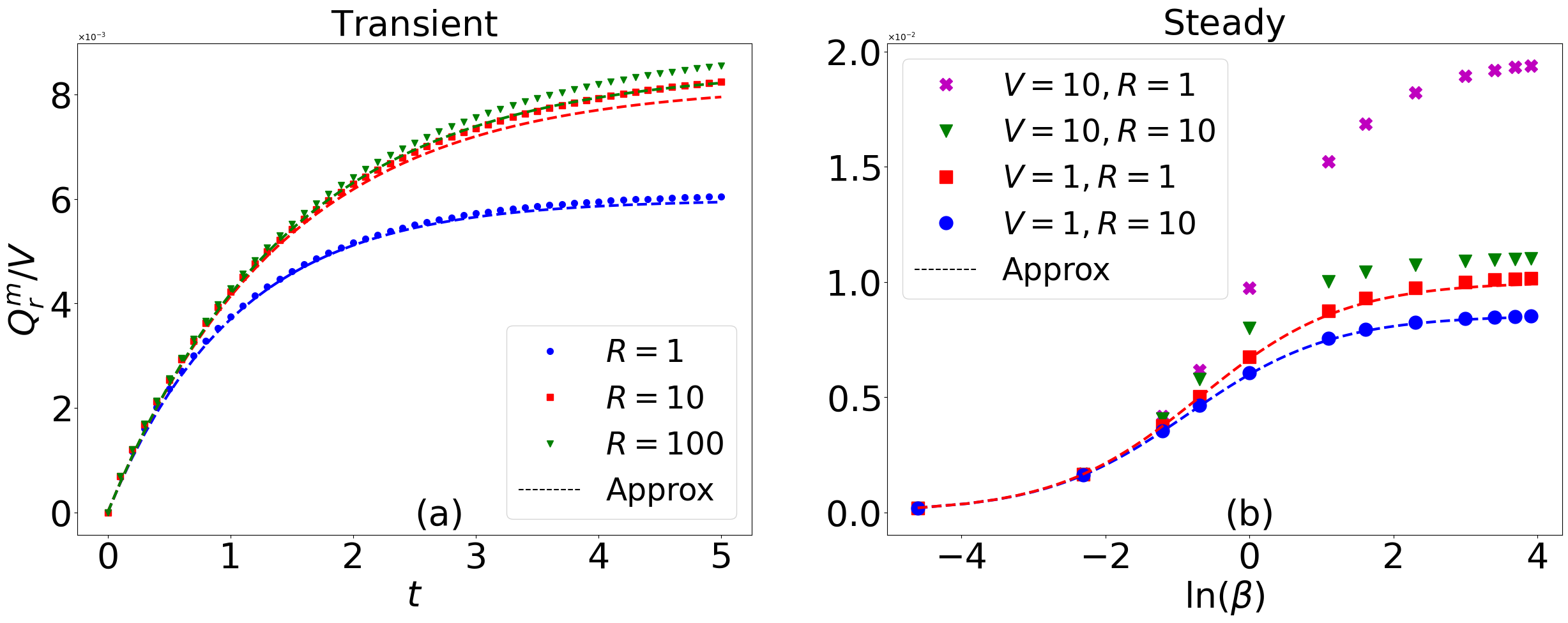}
\caption{Numerical (markers) and asymptotic (dashed line) distributions of transient right-side diffuse charge $Q^m_r$ over time($\beta=1,V=1$) and steady-state $Q^m_r$ over $\beta$ in log scale with imposed voltage $V=1$ and $V=10$. The colors of dashed line have the same legend as the colors of markers.}
\label{fig:qr}
\end{figure*}

\begin{figure}[t]
\centering
\includegraphics[width=\linewidth]{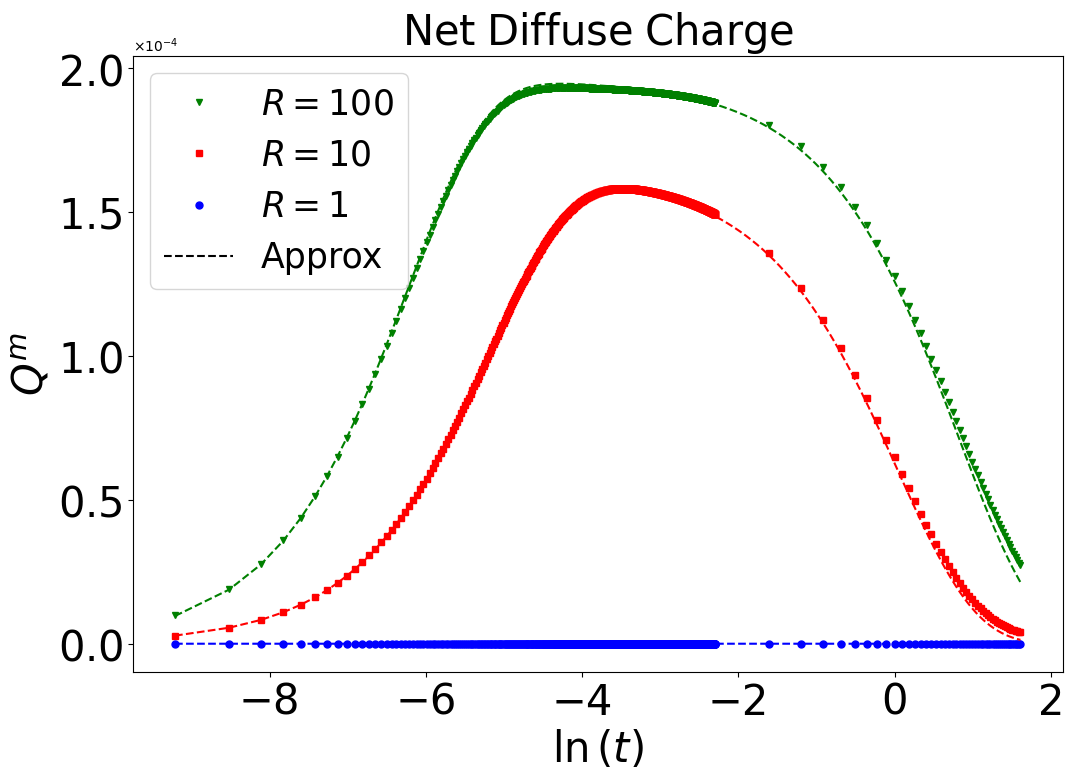}
\caption{Numerical(markers)  and asymptotic(dashed line) distributions of net diffuse charge $Q^m$ over time in log scale with $R=1,10,100$. The colors of dashed line have the same legend as the colors of markers. The parameter setup is $\beta=1$ and $V=1$}
\label{fig:qt}
\end{figure}

\section{Result Discussion}\label{results}

In this section we present results obtained from both the asymptotic approximation \eqref{philc}-\eqref{Qmlc} and numerical solutions of the nonlinear PNP equations  Eqs.~\eqref{NPs}-\eqref{zerofluxe}, \eqref{fv}-\eqref{bcend} with the ``short-time" initial conditions Eq.~\eqref{ICs}-\eqref{ICe}. The numerical solution is performed with Dedalus~\cite{Lecoanet:2020}, a Python package based on spectral method. The default dimensionless parameter setup for the simulations is  $\epsilon = 0.01$, $V = 1$, $\beta=1$, and $\gamma = 10^{-\frac{1}{4}}\;(R = \gamma^{-4} = 10)$. If not specified, parameters in all the subsequent figures use the same values as the default setup. The results for $R=0.1$ are omitted because they are obtained from R=10 by replacing $\gamma$ with $\gamma^{-1}$ and their profiles are symmetric to the profiles with $R=10$ w.r.t the origin. If interested only in the membrane-charging dynamics, the numerical solution of Eq.~\eqref{NPs}-\eqref{robinbc}, \eqref{zerofluxe}-\eqref{bcend} should be used with the ``long-time" initial conditions Eq.~\eqref{longICs}-\eqref{longICe}, instead of Eq.~\eqref{zerofluxs}.

\subsection{Short-time charge-relaxation dynamics }
The numerical results for the short-time evolution of the potential $\phi$, charge density $\rho$ and concentration variation $\Delta c$ are shown in Fig.~\ref{fig:CR}, respectively. Concentration variation $\Delta c$ is the difference between concentration and bulk concentration, i.e., $\Delta c_l = c_l-\gamma^2$ and $\Delta c_r = c_r - \gamma^{-2}$. The asymptotic $\phi$ matches well with the numerical PNP solution and both exhibit the fast charge-relaxation dynamics in the $O(\epsilon)$ time range. Notably, the potential remains nearly continuous across the membrane and the Debye layers in all three figures, with visible changes only starting to occur after $t=0.01$. This demonstrates that the dimensionless time scale of the short-time dynamics is indeed $O(\epsilon)$. However, the inner layers of charge density $\rho$ and depletion of concentration $c$ near the membrane indicate a small deviation from the asymptotic result in the short-time range which appears to be lower order.

\subsection{Membrane-charging dynamics}

\subsubsection{Electric potential and ion distributions}
The numerical results depicting membrane-charging transient states of $\phi$, $\rho$ and $\Delta c$ distributions during membrane-charging regime are presented in Fig.~\ref{fig:MC}, respectively. The composite asymptotic approximations are generally indistinguishable from the numerical results. As shown in Fig.~\ref{fig:MC}, at the membrane the potential displays initial continuity followed by a growing  jump, and the bulk potential approaches a constant at steady state. The Debye layers develop on a  $O(1)$ scale. Even though the charge density profile is asymmetric relative to the
membrane, the net diffuse charge on the membrane remains zero at leading order. 

The electric potential $\phi$ and charge $\rho$ profiles reveal three regions along $x$: a neutral outer region and two charged inner layers of $O(\epsilon)$ width on either side of the membrane. The asymptotic approximations for $\phi$ and $\rho$ agree well with the numerical results. However, the concentration distribution for $\Delta c$ in Fig.~\ref{fig:MC} highlights a  discrepancy in the leading-order asymptotic approximation of the concentration $c$ compared to the other distributions. While the composite asymptotic solutions accurately capture the behaviors of the Debye layer and bulk region\cite{Bazant:2004}, there are some small differences in the intermediate areas within the bulk region. However, these errors in the approximation of the concentration $c$ remain within the order of error tolerated by the asymptotics presented in Fig. \ref{fig:MC}, up to $7.47\times 10^{-4}$($<\epsilon$).

\subsubsection{Transient and steady state transmembrane potential and diffuse charge}\label{transmembrane}
The evolution and steady state of the transmembrane potential $V_m$ and the right-side diffuse charge $Q^m_r$ for different membrane capacitance $\beta$ and initial concentration ratio $R$ are shown in Fig.~\ref{fig:vm} and Fig.~\ref{fig:qr}. Since $Q^m_l + Q^m_r = 0$, it is sufficient to only present the profile of $Q^m_r$. Additionally, from Eq.~\eqref{qmv} it is evident that $Q^m_r$ is a linear function of $V_m$ for fixed $\beta$, so a plot of $Q^m_r$ as a function of $R$ is omitted as the behavior mirrors the $V_m$ plot. The evolution plots show that both $V_m$ and $Q_r$ rise from zero monotonically to reach the steady state. A larger capacitance $\beta$ and a higher ratio of initial concentration $R$ contribute to a longer relaxation time. $V_m$ decreases as $\beta$ increases, while the diffuse charge $Q^m_r$ increases. Conversely, both transmembrane potential and diffuse charge $Q_r$ decrease with  increasing $R$. 

With a small capacitance, the membrane stores little charge and the voltage difference imposed at the electrode occurs entirely across the membrane, similar to the predictions of the leaky dielectric model. Conversely, with a large capacitance, the membrane stores more charge, and the transmembrane potential decreases to zero, resulting in an early continuous potential. Both $V_m$ and $Q^m_r$ exhibit symmetric distributions over log values of $R$, as expected from the symmetric nature of scaling. The transmembrane potential and diffuse charge decrease with increasing asymmetry in the electrolyte concentration. Fig.~\ref{fig:vm} and Fig.~\ref{fig:qr} also suggest that a larger imposed voltage $V$ results in a higher deviation from the asymptotic result for both $V_m$ and $Q^m_r$, as nonlinear effects become apparent with increasing voltage $V$.

\subsubsection{Net diffuse charge on the membrane}
Fig.~\ref{fig:qt} shows that the composite asymptotic approximations of net diffuse charge Eq.~\eqref{totaldiffusecharge} match well with the numerical results. For $R\neq1$, a small amount of charge of order $O(\epsilon^2)$ is accumulated inside Debye layer during the short-time $T_{cr}$, and then dissipated during the long-time $T_{mc}$. The net diffuse charge in the combined Debye layers remains zero at the leading order but the asymmetry of conductivity causes a transient accumulation and relaxation of charge at order $O(\epsilon^2)$, due to the imbalance of the bulk currents in the short-time $T_{cr}$. The symmetry $R=1$ case has zero net diffuse charge because the bulk currents are balanced all the time.

\section{Conclusion}
\label{conclusion}
We have analyzed the electrokinetics across a zero-thickness, non-conducting membrane separating  electrolyte solutions with different ion concentrations. The asymptotic approximations agree well with the numerical results. In the limit of $\frac{\beta}{\epsilon}\gg 1$, the asymmetry in conductivity gives rise to a ``short-time" charge-relaxation dynamics, accompanied by appearance of net diffuse charge in the Debye layers, followed by a ``long-time" dynamics where the net diffuse charge near the membrane dissipates. The numerical and asymptotic results show the ``short" and ``long" time scales are sensitive to the conductivity mismatch. The asymmetry slows the membrane charging, increases the net diffuse charge in the membrane and decreases the transmembrane potential. Our analysis formally derives that the interfacial capacitance in the leaky dielectric model is equivalent to a geometric average of the membrane and adjacent Debye layers.
\section{Acknowledgments}
This research was supported by NIGMS award 1R01GM140461.

\section{Data availability}
The data that support the findings of this article are openly available \cite{??}
\appendix
\section{Formal asymptotics of short-time charge-relaxation dynamics}\label{appini}
After rescaling the inner time coordinate $T = \frac{t}{\epsilon}$, the PNP equations become
\begin{subequations}
\begin{align}
    & \frac{\partial \rho}{\partial T} = \epsilon^2\left(\frac{\partial^2 \rho}{\partial x^2} + \frac{\partial}{\partial x}\left(c\frac{\partial \phi}{\partial x}\right)\right)\,,\label{spnps}\\
    &\frac{\partial c}{\partial T} = \epsilon^2\left(\frac{\partial^2 c}{\partial x^2} + \frac{\partial}{\partial x}\left(\rho\frac{\partial \phi}{\partial x}\right)\right)\,,\\
    &-\epsilon^2\frac{\partial^2 \phi}{\partial x^2} = \rho\,.\label{spoi}
\end{align}
\end{subequations}
The Poisson equation \eqref{spoi} indicates that the charge density $\rho_{r,l}$ is at most second order. Hence in the outer region we seek a solution as  a regular asymptotic expansion:
\begin{subequations}
\begin{align}
    & c = c^{(0)} + \epsilon c^{(1)} + \dots\,,\label{sexps}\\
    & \rho = \epsilon^2 \rho^{(2)} + \epsilon^3 \rho^{(3)} + \dots\quad (\rho^{(0)} = \rho^{(1)} = 0)\,,\\
    & \phi = \phi^{(0)} + \epsilon \phi^{(1)}+\dots\,.\label{sexpe}
\end{align}
\end{subequations}
From Eqs.~\eqref{ICs}-\eqref{ICe}, the leading order initial conditions are
\begin{subequations}
\begin{align}
    & \phi^{(0)}(T=0) = Vx\,,\\
    & \rho^{(0)}(T=0) = 0\,,\\
    & c^{(0)}(T=0) = \begin{cases}
        \gamma^2 & -1<x<0\,,\\
        \gamma^{-2} & 0<x<1\,.
    \end{cases}
\end{align}
\end{subequations}
Now substituting expansions \eqref{sexps}-\eqref{sexpe} into equations \eqref{spnps}-\eqref{spoi} we have at leading order
\begin{subequations}
\begin{align}
    & \rho_l^{(2)} = \rho_l^{(2)}(T=0,x)e^{-\gamma^2 T} = 0\,,\label{appbulks}\\
    & \rho_r^{(2)} = \rho_r^{(2)}(T=0,x)e^{-\frac{T}{\gamma^2}} = 0\,,\label{appbulkm}\\
    \Rightarrow & \rho^{(2)} = 0\,,\\
    & c^{(0)} = \begin{cases}
        \gamma^2 & -1<x<0\,,\\
        \gamma^{-2} & 0<x<1\,.
    \end{cases}\label{appbulkk}
\end{align}
\end{subequations}
Thus, from Eqs.~\eqref{spoi} and \eqref{appbulkk}, the leading-order bulk potentials are linear:
\begin{subequations}
\begin{align}
    & \phi^{(0)} = \begin{cases}
        -V+E_l(T)(x+1) & -1<x<0\,,\\
        V+E_r(T)(x-1) & 0<x<1\,.
    \end{cases}
\end{align}
\end{subequations}
Now we turn to the inner regions. Since $\epsilon$ represents the dimensionless Debye length, we chose a inner space coordinate $Y = \frac{x}{\epsilon}$. Accordingly, the original PNP equations \eqref{spnps}-\eqref{spoi} transform into
\begin{subequations}
\begin{align}
    &\frac{\partial P}{\partial T} = \frac{\partial^2 P}{\partial Y^2} + \frac{\partial}{\partial Y}\left(C\frac{\partial \Phi}{\partial Y}\right)\,,\\
    &\frac{\partial C}{\partial T} = \frac{\partial^2 C}{\partial Y^2} + \frac{\partial}{\partial Y}\left(P\frac{\partial \Phi}{\partial Y}\right)\,,\\
    &-\frac{\partial^2 \Phi}{\partial Y^2} = P\,.
\end{align}
\end{subequations}
Variables $C$, $P$ and $\Phi$ are used to denote concentration, density and potential inside the Debye layers. Applying a regular inner expansion 
\begin{subequations}
\begin{align}
    & P = P^{(0)}+\epsilon P^{(1)}+\cdots\,,\\
    & C = C^{(0)}+\epsilon C^{(1)}+\cdots\,,\\
    & \Phi = \Phi^{(0)}+\epsilon\Phi^{(1)}+\cdots\,.
\end{align}
\end{subequations}
Solution  $P_{r,l0} = 0,C_{r,l0} = \mathrm{constant}, \Phi_{r,l0} = \mathrm{constant}$ satisfies the leading-order inner equation and is proved using weak-field linearization, so the leading-order inner solutions are
\begin{subequations}
\begin{align}
    & P^{(0)} = 0\,,\\
    & C^{(0)} = \begin{cases}
        \gamma^2 & Y<0\,,\\
        \gamma^{-2} & Y>0\,,
    \end{cases}\\
    & \Phi^{(0)} = \begin{cases}
        -V+E_l(T) & Y<0\,,\\
        V-E_r(T) & Y>0\,.
    \end{cases}
\end{align}
\end{subequations}
Then the Robin \eqref{robinbc} boundary condition shows $V_m^{(0)} = 0$:
\begin{align}
    &-V+E_l(T) = V-E_r(T)\,.\label{appvm}
\end{align}
For the second-order inner equations, we want to keep the nonlinearity. With substitution of the leading-order solution, the second-order inner equations are
\begin{subequations}
\begin{align}
    &\frac{\partial P^{(1)}}{\partial T} = \frac{\partial^2 P^{(1)}}{\partial Y^2} + C^{(0)}\frac{\partial^2 \Phi^{(1)}}{\partial Y^2}\,,\\
    &\frac{\partial C^{(1)}}{\partial T} = \frac{\partial^2 C^{(1)}}{\partial Y^2}\,,\\
    &-\frac{\partial^2 \Phi^{(1)}}{\partial Y^2} = P^{(1)}\,.\label{fpoi}
\end{align}
\end{subequations}
We then take a Laplace transform with respect to the inner time variable $T$. For simplicity, denote the Laplace transform using tilde over the variable: $\Tilde{f} = \Laplace\{f\}$,
\begin{subequations}
\begin{align}
    &s\Tilde{P}_l^{(1)} = \frac{\partial^2 \Tilde{P}_l^{(1)}}{\partial Y^2}-\gamma^2\Tilde{P}_l^{(1)}\,,\label{lap1}\\
    &s\Tilde{P}_r^{(1)} = \frac{\partial^2 \Tilde{P}_r^{(1)}}{\partial Y^2}-\gamma^{-2}\Tilde{P}_r^{(1)}\,.\label{lap2}
\end{align}
\end{subequations}
Solving Eqs.~\eqref{lap1}-\eqref{lap2} and substituting matching conditions $\Tilde{P}_l^{(1)}(-\infty) = \Tilde{P}_r^{(1)}(\infty) = 0$ lead to
\begin{subequations}
\begin{align}
    & \Tilde{P}_l^{(1)} = \Tilde{A}_l(s)e^{\sqrt{s+\gamma^2}Y}\,,\\
    & \Tilde{P}_r^{(1)} = \Tilde{A}_r(s)e^{-\sqrt{s+\gamma^{-2}}Y}\,.
\end{align}
\end{subequations}
Then substituting into the Poisson equation \eqref{fpoi} and applying matching conditions results in 
\begin{subequations}
\begin{align}
    & \Tilde{\Phi}_l^{(1)} = -\frac{\Tilde{A}_l(s)}{s+\gamma^2}e^{\sqrt{s+\gamma^2}Y}+\Tilde{E}_l(s)Y\,,\\
    & \Tilde{\Phi}_r^{(1)} = -\frac{\Tilde{A}_r(s)}{s+\gamma^{-2}}e^{-\sqrt{s+\gamma^{-2}}Y}+\Tilde{E}_r(s)Y\,.
\end{align}
\end{subequations}
Afterwards, substituting the zero-flux \eqref{zerofluxs} and Robin \eqref{robinbc} boundary conditions \eqref{zerofluxs} at $Y=0$ we have
\begin{subequations}
\begin{align}
    & 0=\frac{\partial \Tilde{P}_l}{\partial Y}+\gamma^2\frac{\partial \Tilde{\Phi}_l}{\partial Y} = \frac{s}{\sqrt{s+\gamma^2}}\Tilde{A}_l+\gamma^2\Tilde{E}_l\,,\label{las}\\
    & 0=\frac{\partial \Tilde{P}_r}{\partial Y}+\gamma^{-2}\frac{\partial \Tilde{\Phi}_r}{\partial Y} = -\frac{s}{\sqrt{s+\gamma^{-2}}}\Tilde{A}_r+\gamma^{-2}\Tilde{E}_r\,,\\
    & -\frac{\Tilde{A}_l}{\sqrt{s+\gamma^2}}+\Tilde{E}_l = \frac{\Tilde{A}_r}{\sqrt{s+\gamma^{-2}}}+\Tilde{E}_r\,.\label{lae}
\end{align}
\end{subequations}
Finally, solving Eqs.~\eqref{appvm} and \eqref{las}-\eqref{lae} gives 
\begin{subequations}
\begin{align}
    &\Tilde{E}_l = \frac{2V(\gamma^{-2}+s)}{s(\gamma^2+\gamma^{-2}+2s)}\,,\\
    &\Tilde{E}_r = \frac{2V(\gamma^2+s)}{s(\gamma^2+\gamma^{-2}+2s)}\,.
\end{align}
\end{subequations}
After Laplace inverse transform we have
\begin{subequations}
\begin{align}
    & E_l = \frac{2V}{1+\gamma^4}+V\frac{\gamma^4-1}{\gamma^4+1}e^{-\frac{\left(\gamma^2+\gamma^{-2}\right)T}{2}}\,,\label{appres}\\
    & E_r = \frac{2\gamma^4V}{1+\gamma^4}+V\frac{1-\gamma^4}{\gamma^4+1}e^{-\frac{\left(\gamma^2+\gamma^{-2}\right)T}{2}}\,.\label{appree}
\end{align}
\end{subequations}
Eqs.~\eqref{appres}-\eqref{appree} match the results \eqref{sds}-\eqref{sde} in section \ref{inisec}.

Furthermore, the short-time net diffuse charge is
\begin{align}
\begin{split}
    \Tilde{Q}^m(s) &=\epsilon^2\int_{-\infty}^0\Tilde{P}_l^{(1)}dY+\epsilon^2\int^{\infty}_0\Tilde{P}_r^{(1)}dY\,,
\end{split}\\
\begin{split}
    \Rightarrow Q^m(T) &= 2\epsilon^2 V\frac{1-\gamma^4}{1+\gamma^4}\left(1-e^{-\frac{\left(\gamma^2+\gamma^{-2}\right)T}{2}}\right)\,. 
\end{split}
\end{align}

\section{Composite solutions for the current and net diffuse charge}\label{appcurrent}
As defined in Eq.~\eqref{zerofluxs}, the dimensionless current $J$ is
\begin{align}
    & J = \frac{\partial \rho}{\partial x} + c\frac{\partial \phi}{\partial x}\,.\label{jdef}
\end{align}
The leading-order composite solution of current $Jc$ is the sum of short-time current $J^{(0)}(x,T)$, bulk current $J^{(0)}(x,t)$ and inner current $J^{(0)}(Y,t)$, excluding the overlap part:
\begin{align}
    & J_c = J^{(0)}(x,T)+J^{(0)}(x,t)+J^{(0)}(Y,t)-\mathrm{overlap}\,.
\end{align}
The short-time current and the bulk current come from substituting the solutions of potential, charge density and concentration Eqs.~\eqref{bulks}-\eqref{bulke} and \eqref{outer}-\eqref{outerphi} into Eq.~\eqref{jdef}:
\begin{subequations}
\begin{align}
    & J^{(0)}(x,T) = \begin{cases}
        \gamma^2 E_l(T) & -1<x<0\,,\\
        \gamma^{-2}E_r(T) & 0<x<1\,,
    \end{cases}\label{appjss}\\
    & J^{(0)}(x,t) = j(t)\,.\label{appjb}
\end{align}
\end{subequations}
For the inner current $J^{(0)}(Y,t)$, however, the leading-order inner PNP equation \eqref{mi1}-\eqref{mipo} indicates the zeroth-order $O(1)$ inner current is zero $J^{(0)}(Y,t)=0$, so the leading-order term is the second-order $O(\epsilon)$ inner current $J^{(1)}(Y,t)$. Taking the second-order terms from the inner PNP equation \eqref{inners} leads to
\begin{align}
    &\frac{\partial P^{(0)}}{\partial t} = \frac{\partial J^{(1)}}{\partial Y}\,\label{appjs},
\end{align}
with boundary conditions from the zero-flux condition \eqref{zerofluxs} and matching to the outer:
\begin{subequations}
\begin{align}
    & J^{(1)}(Y=0,t) = 0\,,\\
    & J_l^{(1)}(Y\rightarrow-\infty, t) = J_l^{(0)}(x\rightarrow 0,t)\,,\\
    & J_r^{(1)}(Y\rightarrow\infty, t) = J_r^{(0)}(x\rightarrow 0,t)\,.\label{appje}
\end{align}
\end{subequations}
Now we solve Eqs.~\eqref{appjs}-\eqref{appje} with the substitution of Eqs.~\eqref{mis}, \eqref{innerphi}-\eqref{innerphi2} and \eqref{tt1}-\eqref{tt2}. Then we combine $J^{(1)}$ with Eqs.~\eqref{appjss}-\eqref{appjb} to get the composite solutions $J_c$:
\begin{align}
    & J_c = \begin{cases}
    j(t)\left(1-e^{\gamma \frac{x}{\epsilon}}\right)+\gamma^2e(T) & -1<x<0\,,\\
    j(t)\left(1-e^{-\frac{x}{\gamma\epsilon}}\right)+\gamma^{-2}e(T) & 0<x<1\,.
\end{cases}
\end{align}
For the long-time inner net diffuse charge $Q^m(Y,t)$, we return to Eq.~\eqref{chargederivative}:
\begin{subequations}
\begin{align}
    & \frac{dQ_l}{dt} = -\lim_{x\rightarrow 0}J_l(x,t)\,,\\
    & \frac{dQ_r}{dt} = \lim_{x\rightarrow 0}J_r(x,t)\,,\\
    \Rightarrow &\frac{dQ^m(Y,t)}{dt} = \frac{d\left(Q_r+Q_l\right)}{dt} = \lim_{x\rightarrow 0}\left(J_r(x,t)-J_l(x,t)\right)\,.\label{qmt}
\end{align}
\end{subequations}
Eqs.~\eqref{jrl} and Eqs.~\eqref{neutral} have shown $Q^m(Y,t)$ is zero at the order of $O(1)$. For the order of $O(\epsilon)$, we consider the second order of Eq.~\eqref{qmt}:
\begin{align}
    & \frac{dQ^{m(1)}(Y,t)}{dt} = \lim_{x\rightarrow 0}\left(J_r^{(1)}(x,t)-J_l^{(1)}(x,t)\right)\,.
\end{align}
The second order of Eq.~\eqref{ele} yields
\begin{align}
    &\lim_{x\rightarrow 0}\left(J_r^{(1)}(x,t)-J_l^{(1)}(x,t)\right) = \lim_{x\rightarrow 0}\left(\frac{\partial^2 \phi_l^{(0)}}{\partial t\partial x}-\frac{\partial^2 \phi_r^{(0)}}{\partial t\partial x} \right)\,.
\end{align}
Therefore we have
\begin{subequations}\label{qmtdiff}
\begin{align}
    & \frac{dQ^{m(1)}(Y,t)}{dt} = \left(\gamma^{-2}-\gamma^2\right)\frac{d}{dt}j(t),\\
    & Q^m(Y,t=0) = Q^m(T\rightarrow\infty).
\end{align}
\end{subequations}
The solution of Eqs.~\eqref{qmtdiff} shows the long-time net diffuse charge is
\begin{align}
    & Q^m(x,t) = \epsilon^2 Q^{m(1)}(Y,t) = 2\epsilon^2 V\frac{1-\gamma^4}{1+\gamma^4}e^{-\frac{t}{\tau_m}}.\\ 
\end{align}
Finally, the composite solution of net diffuse charge is
\begin{align}
\begin{split}
    Q^m_c & = Q^m(x,T)+Q^m(x,t)-\mathrm{overlap}\\
    & = 2\epsilon^2 V\frac{1-\gamma^4}{1+\gamma^4}\left(e^{-\frac{t}{\tau_m}}-e^{-\frac{\left(\gamma^2+\gamma^{-2}\right)T}{2}}\right).
\end{split}
\end{align}

\bibliographystyle{unsrtnat}
\bibliography{bibliography.bib}

\end{document}